\documentclass{PoS}

\title{X-ray behaviour of GRBs detected by \emph{INTEGRAL}/JEM-X}

\ShortTitle{X-ray emission from GRBs}

\author{\speaker{A. Martin-Carrillo} and L. Hanlon \\
        Space Science Group, School of Physics, University College Dublin, Belfield, Dublin 4, Ireland\\
        E-mail: \email{antonio.martin-carrillo@ucd.ie}}

\abstract{\emph{INTEGRAL}'s JEM-X instrument offers a very rare opportunity to observe the full prompt X-ray emission from GRBs and the transition to the afterglow phase. A study of prompt X-ray flares in some bursts from the \emph{INTEGRAL} GRB sample and the early X-ray post-GRB emission from 3--35\,keV is presented here. Significant post-GRB emission above 10\,keV is observed for GRB\,041219A and GRB\,081003A.}

\FullConference{"An INTEGRAL view of the high-energy sky (the first 10 years)"
9th INTEGRAL Workshop and celebration of the 10th anniversary of the launch,\\
		October 15-19, 2012\\
		Bibliotheque Nationale de France, Paris, France}

\begin{document}

\section{Introduction}
Gamma-ray bursts (GRBs) can be classified based on their durations as long GRBs ($>$\,2\,s) and short GRBs ($<$\,2\,s). However, the duration of a GRB, typically measured by the T$_{90}$ duration, strongly depends on the sensitivity of the instrument and the energy range over which the burst has been detected \cite{sakamoto}. GRB pulses tend to be wider at lower energies \cite{fenimore}, so their durations tend to be considerably longer at lower energies than at higher energies.

The launch of \emph{Swift} brought a major improvement in the study of early X-ray afterglows of GRBs thanks to its fast and autonomous re-pointing of the X-ray (and optical) instrument on board. The typical delay of the \emph{Swift} X-ray observations is $\sim$70\,s, and thus, for long GRBs it is possible to obtain late prompt emission simultaneously in the X-ray and $\gamma$-ray bands. However, due to the very small field of view of the XRT instrument, X-ray observations of the whole prompt emission phase are highly unlikely. 

Since its launch, \emph{INTEGRAL} has observed 22 GRBs (out of $\sim$\,90 GRBs localised by IBIS) with an off-axis angle equal to or smaller than $\sim$\,5$^{\circ}$. GRBs at such small off-axis angles can be observed, during their prompt emission, by the JEM-X (3--35\,keV) instrument on board \emph{INTEGRAL}, providing additional spectral information for the prompt emission and the transition to the afterglow phase. The availability of simultaneous X-ray data for some \emph{INTEGRAL} bursts motivated a search for X-ray afterglow emission in the sample of 22 GRBs detected by JEM-X. Post-GRB emission in this context is defined as emission occurring after the T$_{90}$ burst duration in the $\gamma$--ray band (20--200\,keV). A rapid decline in the X-ray light curve is expected after the prompt emission, based on the canonical X-ray afterglow behaviour, with the possibility of X-ray flares at times $\sim$T$_{0}$+100--1000\,s, where T$_{0}$ is the trigger time \cite{zhang}. The total energy in the early X-ray flares can, in some cases, be comparable to the burst itself \cite{falcone}. These flares are currently associated with internal shocks (as in the prompt phase) rather than the external shocks associated with the afterglow emission.

X-ray afterglows are typically observed in the 0.3--10\,keV energy range (\emph{XMM-Newton}, \emph{Chandra}, and \emph{Swift}). To date, few GRB afterglows have been seen at energies higher than 10\,keV. Emission up to 60\,keV was seen, for the first time, in GRB\,990123 using BeppoSAX \cite{maiorano}. \emph{INTEGRAL} has previously detected hard X-ray afterglow emission, up to 100\,keV, with its instrument IBIS/ISGRI, for GRB\,060428C for about 20\,s after the prompt emission had finished \cite{grebenev}. Thus, JEM-X's sensitivity up to 35\,keV provides a window to study the afterglow emission at energies higher than 10\,keV and could be used to close the energy gap between \emph{Swift}/XRT (0.3--10\,keV) and IBIS/ISGRI (20--1000\,keV) in those cases with simultaneous \emph{Swift} and \emph{INTEGRAL} data.

\section{Data sample}
\begin{figure}[h!]
\centering
\includegraphics[angle=90,width=9cm]{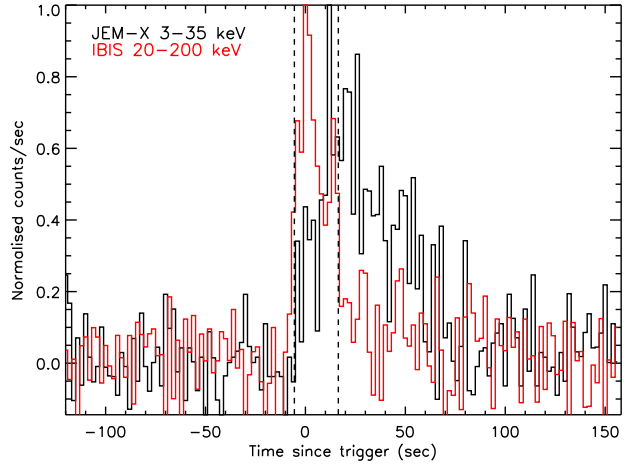}
\caption[]{Gamma-ray and X-ray light curves of GRB\,081003a. The JEM-X (3--35\,keV) light curve is shown in black, while the IBIS (20--200\,keV) light curve is shown in red. The dashed vertical lines represent the T$_{90}$ duration measured by IBIS.}
\label{intsample}
\end{figure}

All 22 GRBs that have been triggered by IBIS/ISGRI within $\sim$\,5$^{\circ}$ are detected by JEM-X. These GRBs are long and in most of the cases weak (peak flux $<$\,0.7\,ph\,cm$^{-2}$\,s$^{-1}$ in the 20--200\,keV energy range). Due to the sensitivity limit of JEM-X, this results in poor quality light curves in most of the 22 cases, with a lot of background noise and low resolution spectra. In order to usefully study the transition between the prompt and afterglow emission, this sample must be further reduced to those cases where good quality observations have been obtained with the JEM-X instrument. Thus, three conditions were imposed on the original sample: a peak flux $>$\,1.5\,ph\,cm$^{-2}$\,s$^{-1}$ in the 3--35\,keV energy range, a fluence $>$\,2.5$\times$\,10$^{-7}$\,erg\,cm$^{-2}$ in the 3--35\,keV band, and X-ray hardness, measured as the fluence ratio between the energy bands 3--10\,keV and 10--35\,keV, to be close to 1. These conservative criteria result in a subsample of 5 GRBs (GRB\,041219A, GRB\,050520, GRB\,051105B, GRB\,081003A, and GRB\,090817). A good example of a GRB with long-lasting emission in its X-ray light curve is shown in Fig.~\ref{intsample}. In this case, GRB\,081003a, the X-ray emission is clearly seen well after the $\gamma$-ray emission (red curve) is finished.

To create X-ray spectra, the X-ray light curves of the GRB candidates for X-ray afterglow emission were re-binned until each point had a signal-to-noise ratio of at least 5 above the background level. The spectrum generated for each interval was then fit by a simple power-law model. The analysis of the data was carried out using \emph{OSA}v9 and \emph{XSPEC}v12.7.

\section{X-ray flares in the late prompt emission}
About 5\% of \emph{Swift}/BAT GRBs have soft pulses at $\sim$\,100--300\,s after the burst trigger, consistent with the times of the X-ray flares commonly observed by \emph{Swift}/XRT. Both the temporal and spectral properties of these soft pulses resemble those of the ``standard" X-ray flares. Thus, if these pulses are attributed to X-ray flares, the T$_{90}$ measurements of this 5\% GRBs would be overestimated.

\begin{figure}[ht]
\centering
\includegraphics[angle=90, width=7.5cm]{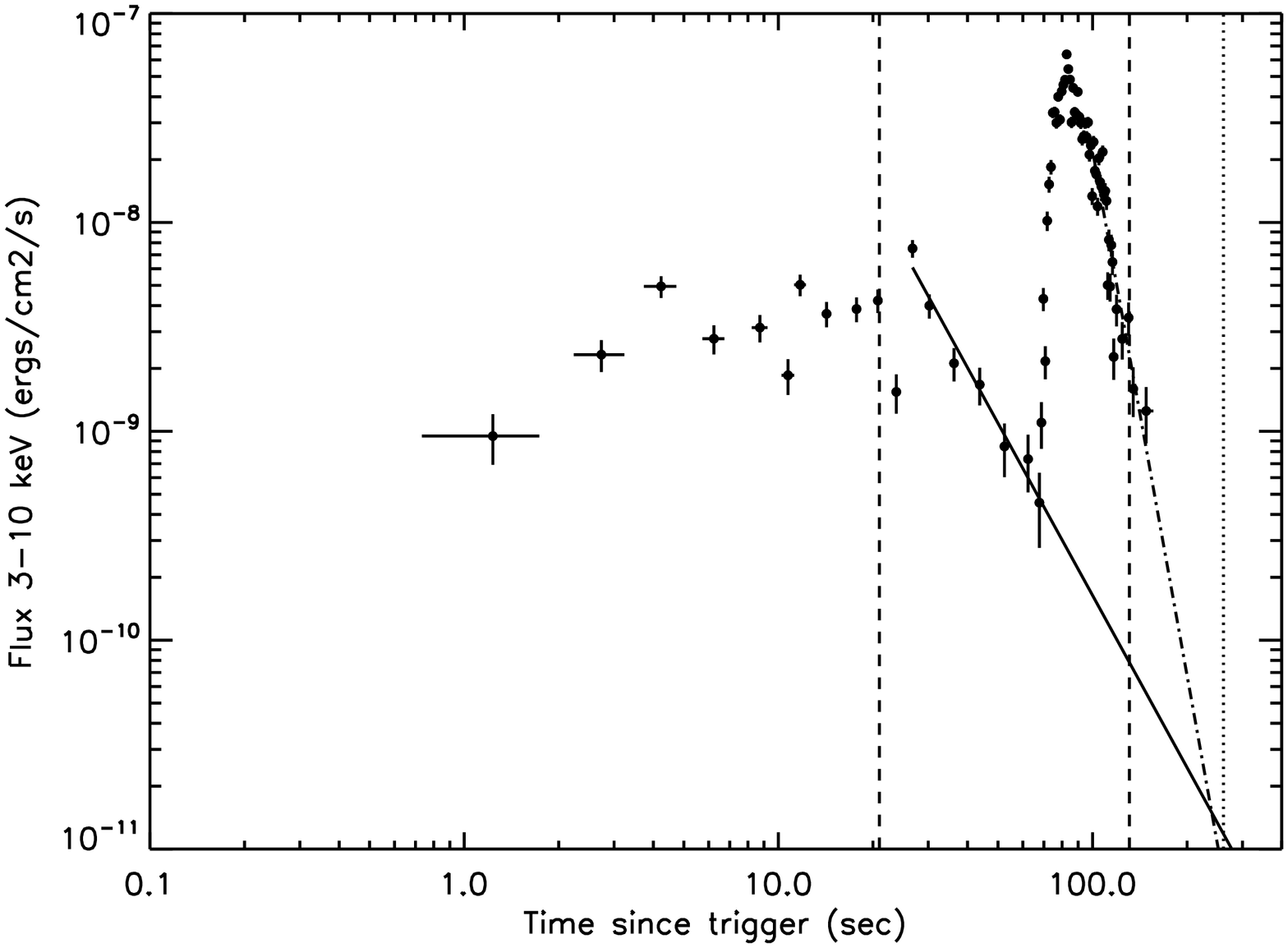}
\includegraphics[angle=90, width=7.5cm]{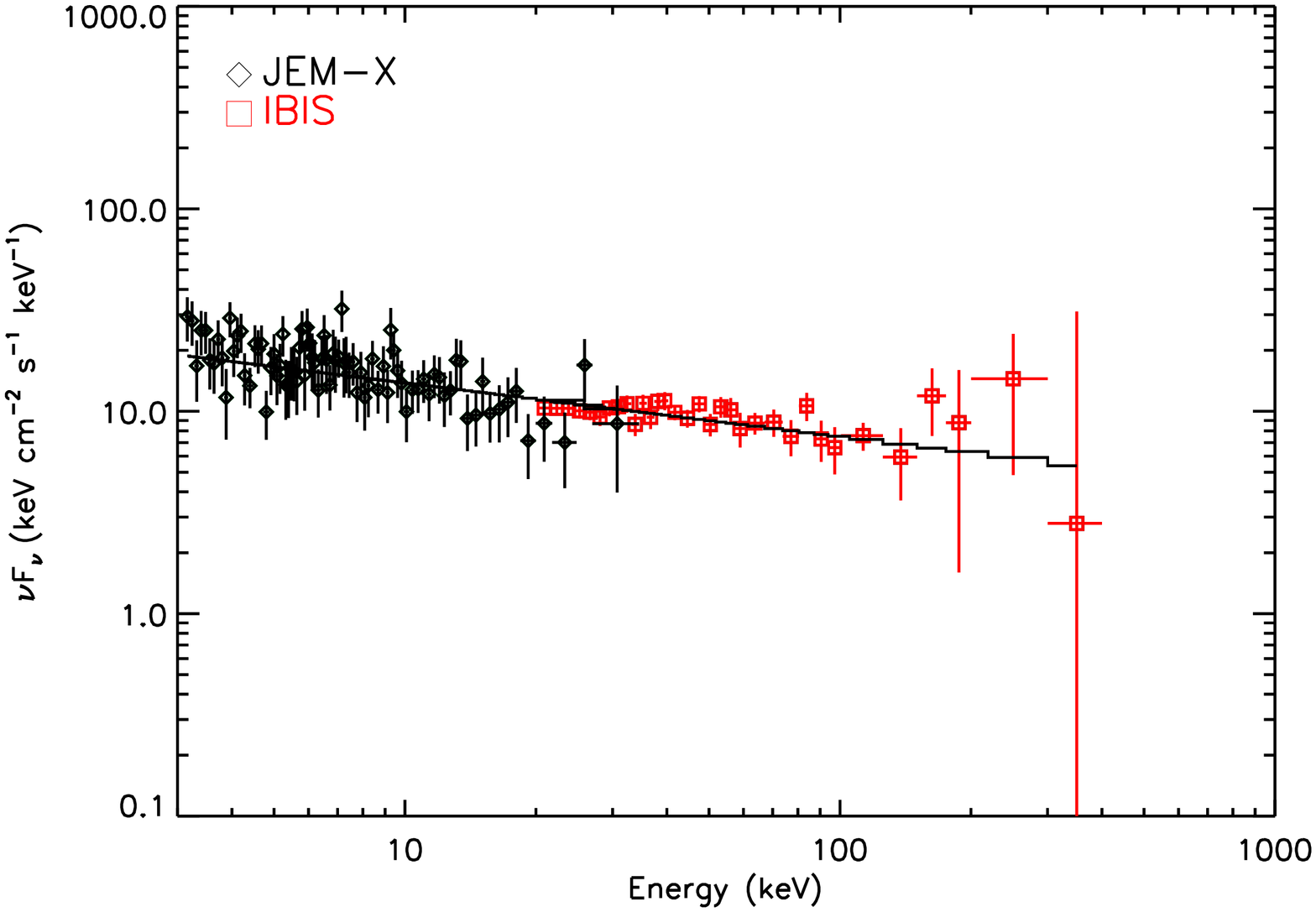}
\caption[]{\textit{Left panel:} JEM-X light curve of the precursor and soft flare (T$_{0}$\,+\,80\,s) observed during the prompt emission of GRB\,041219A, in the 3--10\,keV energy range. The vertical dashed lines represent the end of the precursor and the end of the flare as seen in the 20--200\,keV energy range. The vertical dotted line shows the beginning in $\gamma$-rays of the main pulse of the prompt emission. The solid line corresponds to the best power-law fit to the decay of the precursor, while the dot-dashed line shows the best power-law fit to the decaying phase of the soft flare. \textit{Right panel:} Spectral fit of the combined IBIS and JEM-X data of the soft X-ray flare. The solid line represents the best fit power-law model.}
\label{041219a}
\end{figure}

The \emph{INTEGRAL} GRB sample, shows similar behaviour in $\sim$\,4\% of its bursts. Two of these bursts (GRB\,041219A and GRB\,090817) are also in the JEM-X set of 5 GRBs meeting the selection criteria. Fig.~\ref{041219a} shows the soft X-ray light curve of GRB\,041219A \cite{mcbreen} during the first 200\,s after the trigger i.e. before the main pulse of the prompt emission starts at T$_{0}$\,+\,230\,s (indicated by the vertical dotted line). The first vertical dashed line represents the end of the $\gamma$-ray emission (20--200\,keV) of the precursor. After peaking in X-rays, the precursor light curve declines with a power-law decay of index $\delta$\,=\,-2.74\,$\pm$\,0.21. This value is commonly seen in the early afterglows of GRBs exhibiting strong flares. Therefore, if the soft flare detected at T$_{0}$\,+\,80\,s, is part of the afterglow, its flux should return to the expected value marked by the precursor decay (solid line in Fig.~\ref{041219a}). The beginning of the intense prompt emission (dotted line), limits the length of the good time interval, GTI, which can be used to find this value. It has been typically seen that with GTIs of 50\,s (maximum length allowed in this interval), fluxes $>$\,5\,$\times$\,10$^{-10}$\,ergs\,cm$^{-2}$\,s$^{-1}$ can be detected. No detection is made during the interval after the flare, which indicates that the flux is rapidly decaying with time. The power-law fit of the decaying phase of the soft flare results in a power-law index of $\delta$=\,-8.4\,$\pm$\,0.8 (dashed-dotted line). The projection of this decay with the one measured before the flare is also shown in Fig.~\ref{041219a}. Both precursor and soft flare decays would cross each other at $\sim$\,230\,s after the trigger. The spectrum of the flare (right panel Fig.~\ref{041219a}) is well fit by a single power-law with photon index of $\alpha$\,=\,2.26\,$\pm$\,0.02.

\section{Transition from prompt to afterglow emission}
\begin{figure}[h!]
\centering
\includegraphics[angle=90, width=7.5cm]{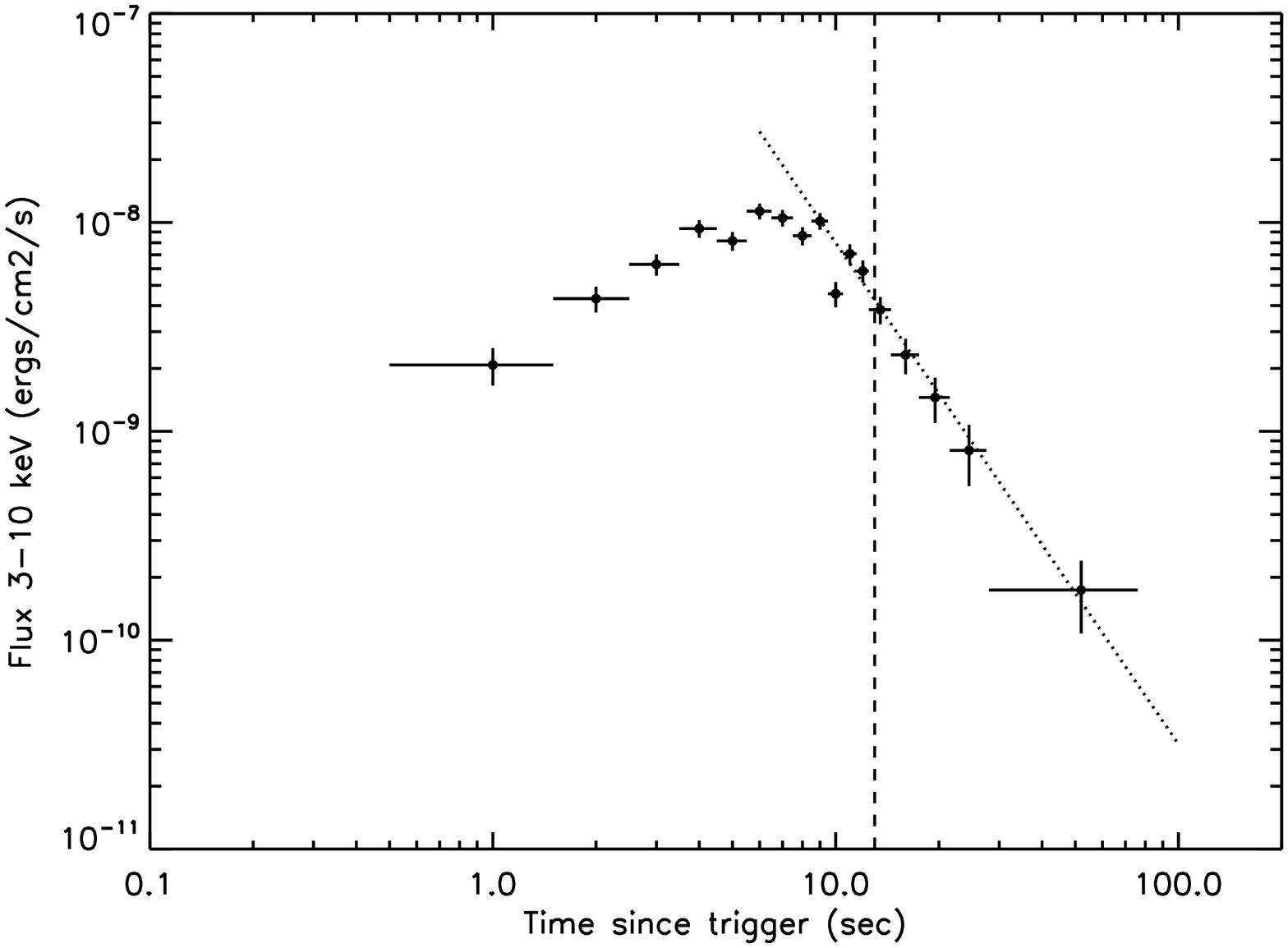}
\includegraphics[angle=90, width=7.5cm]{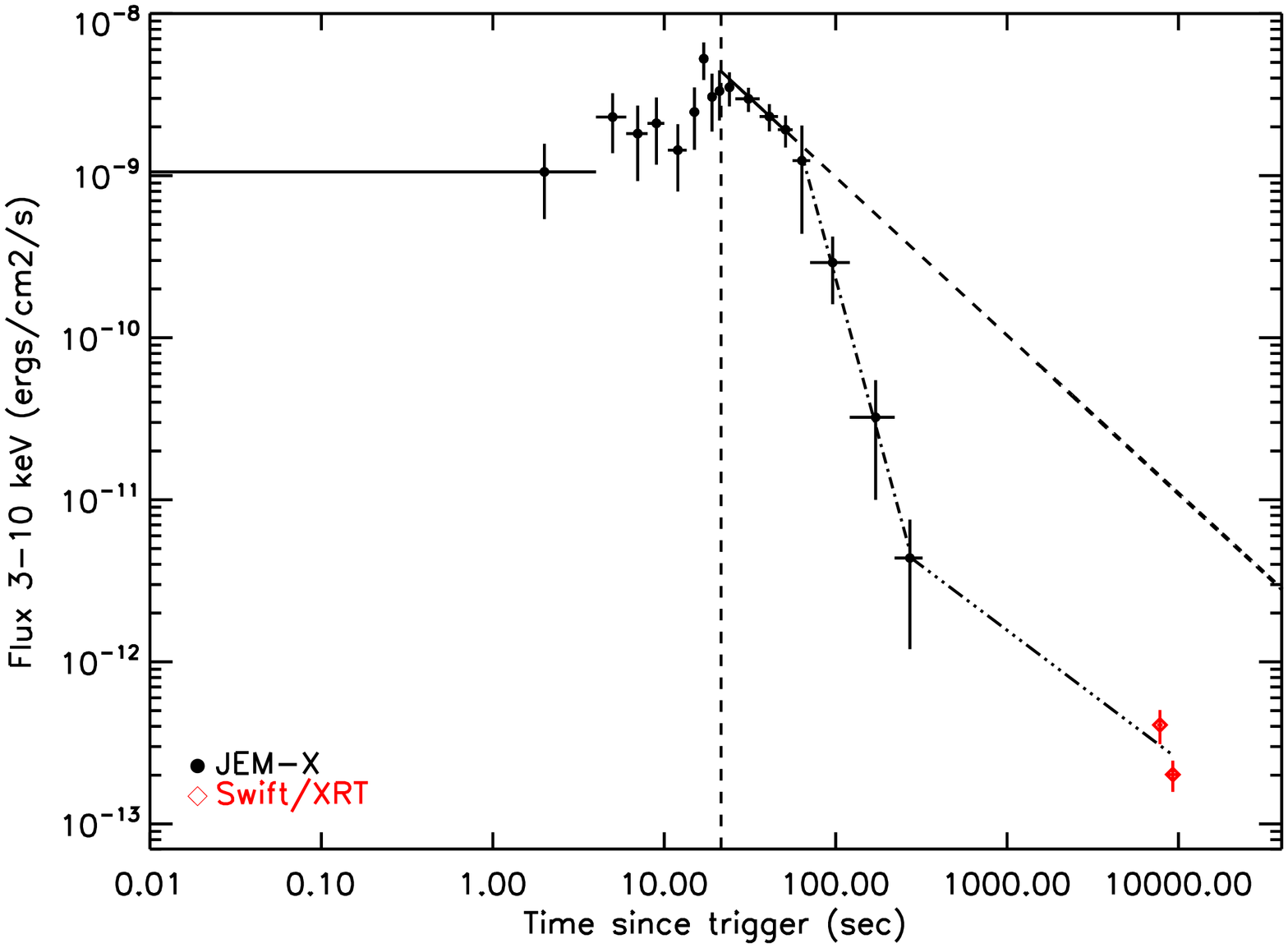}
\caption[]{JEM-X afterglow light curve of GRB\,051105B (left panel) and GRB\,081003A (right panel) in the 3--10\,keV band. The vertical dashed line represents the end of the T$_{90}$ emission (20--200\,keV). The dotted line shows the best power-law fit of the decaying phase of the prompt emission. The Swift/XRT data at $\sim 10^4$\,s are also shown (red diamonds) for GRB\,081003A.}
\label{postgrb}
\end{figure}

Fig.~\ref{postgrb} shows the X-ray GRB emission for two GRBs from the JEM-X sample considered in this analysis (GRB\,051105B and GRB\,081003A, in the left and right panel respectively). As shown in these two examples, the transition and the early afterglow seen in the JEM-X data can be quite different from case to case.

In the case of GRB\,051105B (left panel Fig.~\ref{postgrb}), the X-ray decay differs from the observed temporal shape seen at hard X-rays with IBIS/ISGRI, where the burst ends more abruptly. The X-ray decay phase is observed up to T$_{0}$+50\,s with $\delta$=-2.4\,$\pm$\,0.1. This value agrees with the average decay seen in FRED GRB tails of $\sim$\,2.44 \cite{kocevski}.

GRB\,081003A shows post-GRB emission with different decaying phases (right panel Fig.~\ref{postgrb}). The early post-burst X-ray emission, from T$_{0}$+22\,s to T$_{0}$+70\,s, decays as a power-law of slope -0.98$\pm$0.05. Thereafter the decay slope steepens to a value of -3.88$\pm$0.11 until T$_{0}$+271\,s, consistent with high latitude emission. Later observations (T$_{0}$+10$^{4}$\,s) by \emph{Swift} indicate that the X-ray light curve must flatten out sometime after T$_{0}$+271\,s. A lower limit on the slope of -0.80$\pm$0.14 is derived, assuming the flattening starts at T$_{0}$+250\,s. The early shallow phase lasting $\sim$\,50\,s can be attributed to pure prompt emission, where the pulse at lower energies is wider than the one seen at higher energies. However, the duration between the X-ray and $\gamma$-ray bands is greater than usually seen. The second scenario attributes this phase to afterglow emission with strong long-lasting energy injection from the central engine. At T$_{0}$+70\,s the high latitude emission starts to dominate, however, the fact that another shallow phase seems to be required to accommodate the \emph{Swift} data might imply that the central engine does not switch off completely. 

\section{Post-GRB emission above 10\,keV}
Post-GRB emission above 10\,keV is found in two of the 5 GRBs studied (GRB\,041219A and GRB\,081003A). For GRB\,041219A, X-ray emission is detected at the 3.5$\sigma$ level at T$_{0}$+734$\pm$125\,s (250\,s integration time). A power-law model gives the best fit to the spectrum in this interval, with $\alpha$ = 2.2$\pm$0.7 and a flux of 2.9\,$\times$\,10$^{-10}$\,ergs\,cm$^{-2}$\,s$^{-1}$ in the 3--35\,keV energy range. Most of the detected emission comes from the soft band (3--10\,keV), with 20\% of the measured flux detected in the 10--35\,keV band.

\begin{figure}[h!]
\centering
\includegraphics[angle=0, width=9cm]{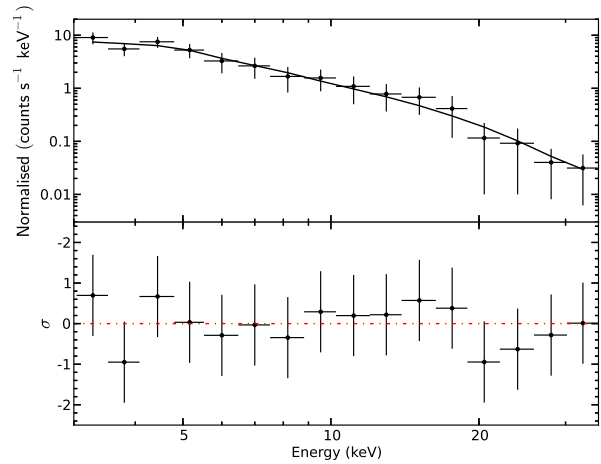}
\caption[]{JEM-X spectrum of GRB\,081003A in the 3--35\,keV energy band from T$_{0}$+22\,s to T$_{0}$+70\,s (shallow phase). The solid line shows the best fit power-law model. The residuals of the fit are shown in the lower panel.}
\label{transition}
\end{figure}

In the case of GRB\,081003A, hard X-ray emission is observed during the shallow decay phase. The spectrum is well fit by a single power-law model with a photon index of $\alpha$ = 2.25$\pm$0.24 with a flux of 7.4\,$\times$\,10$^{-9}$\,ergs\,cm$^{-2}$\,s$^{-1}$ in the 3--35\,keV band (Fig.~\ref{transition}). Around $\sim$\,43\% of this flux comes from radiation above 10\,keV. This percentage of hard X-ray emission decreases sharply to $\sim$\,20\% at T$_{0}$+63.5\,s, right before the start of the high latitude emission, becoming negligible afterwards. 

\section*{Acknowledgements}
This research is based on observations with \emph{INTEGRAL}, an ESA project with instruments and science data centre funded by ESA member states (especially the PI countries: Denmark, France, Germany, Italy, Switzerland, Spain), Poland and with the participation of Russia and the USA. AMC and LH acknowledge support from Science Foundation Ireland under grant 09/RFP/AST/2400.

\end{document}